\documentstyle[PASJadd,epsbox]{PASJ95}

\newcommand{\vol}{--}
\newcommand{\no}{-}
\newcommand{\lett}{}
\newcommand{\spage}{----}
\newcommand{\rdate}{1997 November 13}
\newcommand{\adate}{1999 August 2}

\begin{document}
\setcounter{page}{1}

\title{Possible Young Stellar Objects without Detectable CO Emission}

\author{ Ikuru {\sc Iwata},$^{1,3}$
Shin-ichiro {\sc Okumura},$^2$
and Mamoru {\sc Sait\=o}$^3$
\\[12pt]
$^1${\it Toyama Astronomical Observatory, Toyama Science Museum,
49-4 San-no-kuma, Toyama, Toyama 930-0155}
\\
$^2${\it Okayama Astrophysical Observatory, National Astronomical
 Observatory, Kamogata-cho, Okayama 719-0232}
\\
$^3${\it Department of Astronomy, Faculty of Science, Kyoto
University, Sakyo-ku, Kyoto 606-8502}
\\
{\it E-mail (II): iwata@kusastro.kyoto-u.ac.jp}
}

\abst{
Young stellar objects (YSOs) usually appear in molecular clouds 
as infrared objects associated with a molecular envelope.
Wouterloot and Brand (1989, AAA 50.133.012) searched 
1302 IRAS point sources with reliable fluxes at 25, 60, and 100 
$\mu$m near to the galactic plane for $^{12}$CO($J=1-0$) emission;
1077 sources were detected.
Among their far-infrared sources without detectable CO emission,
we found that at least 18 objects are invisible at optical and 
near-infrared wavelengths. 
The infrared spectral indices between 2.2 $\mu$m and 25 $\mu$m 
correspond to those of class I YSOs, and 
the IRAS colors are similar to those of the usual YSOs.
These peculiar far-infrared objects are 
highly concentrated around the galactic plane and 
the distances are estimated to be $\sim$1 kpc.
Although their distribution is away from molecular clouds,
some of them seem
to be associated with large dark clouds or weak radio sources.
These objects are possible YSOs with low CO abundance in the
envelopes.
}

\kword{Stars: formation --- Infrared: stars --- Circumstellar matter --- Young stellar objects: search}

\maketitle
\thispagestyle{headings}

\section
{Introduction}

Most of stars burstly form in molecular clouds.
Low-mass stars are also born in globules that are dispersely 
distributed in spiral arms (e.g., Yun, Clemens 1994).
Based on the spatial distribution, age distribution, and kinematics of 
T Tauri stars, Feigelson (1996) claims that dispersed
T Tauri stars with various ages consist of components drifting outward
from currently active sites of star formation in molecular
clouds, and components formed in now-dissipated cloudlets in past 
molecular cloud complexes (see Neuh\"auser 1997).

Young stellar objects (YSOs) grow in molecular cores
up to the maximum luminosity phase while showing spectral energy 
distributions (SEDs) with a peak at the far-infrared (FIR) wavelength. 
The gas-to-dust mass ratio in molecular cores is considered to be 
similar to the interstellar value, because the core masses 
derived from the intensities of dust emission (e.g., Walker et al. 1990)
are consistent with those derived from 
CO lines or other molecular lines (e.g., Casoli et al. 1986; 
Benson, Myers 1989; Wouterloot et al. 1989).
In low-mass pre-main sequence stars, such as T Tauri stars, 
however, the envelope masses derived from the intensities
of dust emission are larger than those derived from the CO line
intensities (e.g., Dutrey et al. 1996); it has occurred at some stage 
from YSOs to pre-main sequence stage that CO molecules have been removed
from the envelopes or locked up on grains or more complex
molecules in the envelopes (e.g., Aikawa et al. 1997).

In this paper, we show the presence of 18 bright FIR objects peculiar
in the sense that they are not detectable in the $^{12}$CO line, show
the class I YSOs$'$ SEDs and are located away from molecular clouds.
Section 2 describes the sample. 
In section 3 we consider infrared SED indices of the sample using 
data of IRAS and our near-infrared photometry, and show that the indices
correspond to those of class I YSOs. In section 4 we discuss the
IRAS colors, sky distribution, and luminosities of the sample. 
In section 5 we give a summary.

\section
{Sample}

Wouterloot and Brand (1989) made a complete search for 
the $^{12}$CO($J=1-0$) line on 1302 
IRAS point sources with reliable flux densties at 25, 60, and 
100 $\mu$m (Joint IRAS Science Working Group 1988; IRAS PSC)
and with characteristic FIR colors of YSOs in a zone of $\ell =85^\circ$
to $280^\circ$ and $b=-10^\circ$ to $10^\circ$ (we hereafter call those
WB objects);
their beam size and rms noise level were 43$''$ and 0.2--1.0 K 
for SEST observations, and 21$''$ and 0.7--1.5 K 
for IRAM  observations, respectively.
The velocity ranges of their $^{12}$CO observations are 
220 $\rm km~s^{-1}$ with a central velocity of 60 $\rm km~s^{-1}$ 
for SEST observations,
and 208 $\rm km~s^{-1}$ with central velocities of 
$-50$ to 20 $\rm km~s^{-1}$ for IRAM observations.
They detected CO emission on 1077 point sources
(82.7\%), and no emission on another 223 sources.
Wouterloot et al. (1990) considered that WB objects without
detectable CO emission are extragalactic objects because
of the homogeneous sky distribution in the surveyed region,
in contrast with the distribution of WB objects with CO
emission concentrated around the galactic plane.
We examined the literature and optical images on Schmidt plates for all 
223 WB objects without CO emission,
and found that 132 are associated with optical objects, 
which are early-type stars, planetary nebulae, and 
galaxies. But the remaining 91 WB objects do not have any distinct 
optical counterparts at the positions of the IRAS 
point sources. The results are summarized in table 1.

Iwata et al. (1997) carried out near-infrared imaging observations of 
55 objects among the 91 optically invisible WB objects without 
$^{12}$CO emission at the Okayama Astrophysical Observatory 
(OAO).
As a result, 16 objects were found to be galaxies and possible galaxies,
4 objects are H {\sc ii} regions and a possible H {\sc ii} region,
and 14 objects are stars, 
while 21 objects don't show any counterparts, even in 
the $J, H,$ and $K'$ bands.
The 21 WB objects invisible at optical and
near-infrared wavelengths and not 
detectable in $^{12}$CO($J=1-0$) line were
the sample of this study. We call those ``peculiar FIR'' objects.
Iwata et al. (1997) claimed that these objects are mostly
galactic objects because of the sky distribution along the galactic 
plane and the FIR colors colder than those of IRAS galaxies.

Figure 1 shows a relation of the 60 $\mu$m flux density in IRAS PSC
versus the $^{12}$CO($J=1-0$) line intensity for 21 peculiar FIR objects
and 97 WB objects with
$^{12}$CO emission existing at $\ell=85^\circ$ to $105^\circ$.
The CO intensities are the values of Wouterloot and Brand (1989);
the single-line objects are preferentially adopted. For objects
associated with multi-component CO lines 
we adopted a line if it is several times or more stronger
than other components or it has some active features. 
For the peculiar FIR objects the upper limits are shown. 
We used the 97 WB objects shown by dots in figure 1 as a control sample, 
which are representatives of YSOs at around the maximum luminosity
phase in molecular cores (Wouterloot et al. 1990).
A correlation similar to that of the control sample is shown in
Casoli et al. (1986).
Figure 1 indicates that the peculiar FIR objects have relatively less 
strong 60 $\mu$m flux densities and extremely weak CO intensities
among WB objects. 
 
To ascertain the reality of the peculiar FIR objects, we requested  
the Infrared Processing Analysis Center (IPAC) for co-addition of
the IRAS data. As results of the ADDSCAN-SCANPI programs, 
in which 6 to 21 scan data were coadded for each band
of each object, more reliable flux densities
were obtained for the four bands of all objects, except
for a 12 $\mu$m flux density of IRAS 22044+5451 and an 100 $\mu$m flux
density of IRAS 04375+5016.

Table 2 lists 21 peculiar FIR objects with respect to
the number of Wouterloot and 
Brand (1989), IRAS name, galactic coordinate transformed from
the equatorial coordinate of IRAS point source, flux densities
at 12, 25, 60, and 100 $\mu$m quoted from IRAS PSC at the first
line of each object and those of co-added data at the second line,
three infrared colors calculated using the co-added data, infrared 
luminosity at 1 kpc (an assumed distance, described in subsection 4.2),
and mode of our near-infrared observation.
We computed the luminosity $L_{\rm IR}$ using an equation (Emerson 1988),
\begin{eqnarray}
L_{\rm IR}=0.31d^2 (20.653f_{12}+7.538f_{25}+ \nonumber \\ 
4.578f_{60}  +1.762f_{100})L_\odot,
\end{eqnarray}
where $f_\lambda$ is the co-added flux density in Jy and $d$ is 
a distance in kpc.

\section
{Observations and Results}

The near-infrared imaging observations  
were carried out at OAO on 1995 August and 1996 February 
using Okayama Astrophysical System for Infrared imaging and
Spectroscopy (OASIS) with a detector NICMOS-3 equipped on the 
1.88 m reflector  (Yamashita et al. 1995). The FOV was $4'\times4'$
and the pixel size was $0.\!''97$.
The exposure times were 300 s, 150 s, and 225 s at the $J, H,$ and $K'$
bands, respectively, for 18 objects, and 600 s 
at $J$ or $H$ bands for the other three objects (see table 2). 
The details of the 
observations and data reduction were shown in Iwata et al. (1997). 
Figure 2 shows the optical and near-infrared images of 
$1.\!'2 \times 1.\!'2$ around the 21 objects. 
The optical images were made from Digitized Sky Surveys
produced at the Space Telescope Science Institute 
based on photographic data obtained using the 
Oschin Schmidt Telescope on Palomar Mountain.
At the positions of IRAS point sources which are indicated by the 
position uncertainty ellipses quoted in IRAS PSC, we could not find 
any distinct optical or near-infrared counterpart.

During the observations we observed three photometric standard stars, 
HD 44612, BD +0$^{\circ}$1694, and HD 105601 (Elias et al. 1982); 
their $K$ magnitudes are
7.040, 4.585, and 6.685, respectively.
For a detection limit of $S/N$=10, we obtained
the limiting magnitudes to be $K'=$
14.3 to 15.2 for 225 s exposure, where the range of magnitudes was 
due to the different sky conditions.
The magnitude difference of the OASIS $K'$ band (2.16 $\mu$m)
and $K$ band (2.19 $\mu$m) is less than 0.03 mag in the rms value
for more than 50 stars (Okumura et al. 1999, in preparation).
For a similar photometric system, Wainscoat and Cowie (1992) derived
$K'-K$=0 to 0.08 mag for 18 stars. We assume the OASIS $K'$ magnitude 
to be equivalent to the $K$ magnitude.
The 18 peculiar FIR objects from WB 70 to WB 584 listed in table 2
are less bright than $K=14.3$. 
We thus adopt $K=13.7$ as the upper limit for the 
18 peculiar FIR objects,
assuming an interstellar extinction less than 0.$^m$6 at the $K$ band,
i.e., $A_V<5.6$ (Mathis 1990). 

The $K$ magnitude is transformed to flux 
density $f_K$ in Jy, following Bessell and Brett (1988); i.e., 
\begin{equation}
\log f_K= 0.4(13-K)-2.38.
\end{equation}
Equation (2) yields log$f_K=-2.66$ and 
 log($\nu f$)$_K$=11.47 for $K$=13.7. 
 The upper limit of log($\nu f)_K$=11.47 is
less than log($\nu f)_{25}$  at 25 $\mu$m for all of 
the 18 peculiar FIR objects, which are log$(\nu f)_{25}$=13.08
+log($f_{25})>12.3$ for their values of $f_{25}>$0.2 Jy, 
as shown in table 2.
YSOs are classified by the spectral index, 
$\alpha=-d~{\rm log}(\nu f_\nu)/d~{\rm log}~\nu$, between 2.2 $\mu$m
and 25 $\mu$m. Lada (1987) defined class I YSOs to be objects with 
$\alpha_{2.2,25}>0$, and class II YSOs to be $-2<\alpha_{2.2,25}<0$.
Class I YSOs are usually invisible at optical wavelengths because 
of heavy obscuration by dense molecular clouds, and 
they are considered to be at an early stage of stellar evolution. 
Typical class II YSOs are T Tauri stars.
All of the 18 peculiar FIR objects have positive $\alpha$-indices 
corresponding to class I YSOs. For the remaining three objects
(WB 628, WB 901, and WB 1001), we can not give $\alpha_{2.2,25}$.
Although the exposure times at the $J$ and $H$ bands for these objects were
longer than those for the other 18 objects, we found no counterpart 
within the position uncertainty ellipses of the IRAS point sources.
In the following discussion we include these objects in the 
peculiar FIR objects.

The $\alpha$-indices of all of the peculiar FIR objects are also
positive between $\lambda$=25 $\mu$m and 60 $\mu$m, where
$\alpha_{25,60}=~2.63~{\rm log}(f_{60}/f_ {25})-1$ (see table 2).
But the values of $\alpha _{12,25}=~3.137~{\rm log}(f_{25}/f_{12})-1$ are 
negative for 12 peculiar FIR objects with log$(f_{25}/f_{12})<0.32$
(see table 2). 
This means that more than half of the peculiar FIR objects have 
a second peak at between 2.2 $\mu$m and 25 $\mu$m.

\section
{Discussion}
\subsection
{Features of the Peculiar FIR Objects}

Figures 3 and 4 show IRAS color-color diagrams 
for the peculiar FIR objects, the control sample, which are usual YSOs,
and cirrus. 
The cirrus colors indicated by the open circles were measured 
by Boulanger and Perault (1988) in 8 nearby molecular clouds, 
and the filled circle stands for the FIR color of warm cirrus 
in the H {\sc i} cloud at high galactic latitude (Boulanger et al. 1985). 
Clemens et al. (1991) measured the 
infrared colors of 248 globules and small molecular clouds, and
obtained the mean values and dispersions to be 
log$(f_{100}/f_{60})=0.70\pm0.20$, log$(f_{60}/f_{25})=0.68\pm0.34$, 
and log$(f_{25}/f_{12})=0.01\pm0.30$; these are similar to the 
colors of cirrus described in  Boulanger and Perault (1988). 
Clemens et al. interpreted the infrared emission of globules to be cirrus. 
We find in figures 3 and 4 that the IRAS colors of the peculiar
FIR objects resemble those of the usual YSOs rather than the colors of 
cirrus, except for the warm cirrus. 

Among the visually known WB objects without a detectable $^{12}$CO
line, there are early-type stars, and their number is as many as 
about half of the galaxies, as shown in table 1\@. 
In table 3, such 11 early-type stars with reliable 12 $\mu$m 
flux densities are given. 
The seven stars from B5 to A2 are possibly associated with IRAS 
point sources having $f_{60}\simeq$ 2 Jy to 12 Jy, which are 
similar values to those 
of the peculiar FIR objects. The visual magnitudes are 5 to 9 mag,
and thus the $K$ magnitudes are brighter than 9 mag (Bessell,
Brett 1988). In order to make apparent $K$ magnitudes of these stars
be less than 13 mag due to extinction, the extinction value
should be greater than $A_{\sc v} \simeq 37$ (Mathis 1990), 
which is unusually large interstellar extinction. 
The latitude distributions differ
between these stars and the peculiar FIR objects.
The three B0 or OB stars listed in table 3 are also associated with IRAS
point sources with $f_{60} \simeq 36$ Jy to 56 Jy; the values are
one order of magnitude larger than the typical values of the
peculiar FIR objects. Putting these stars at distances three-times 
further than the present distances, we may obtain $m_{\sc v} \simeq 11.5$
to $13$ and $K>12$. If there is an additional interstellar extinction of
$A_{\sc v} >7^{\rm m}$, we may find no object on these positions at visual
and $K$ bands for detection limits of $m_{\sc v}=19$ and $K = 13$.
Such large interstellar extinctions are only caused by the 
presence of molecular clouds in the line of sight. 
But the peculiar FIR objects do not 
have detectable CO emission. We rule out the possibility 
that the peculiar FIR objects mostly correspond to obscured 
early-type stars with an FIR emitting envelope. 

Galaxies have IRAS colors similar to those of YSOs in a range of 
${\rm log}(f_{60}/f_{25})= 0.6 - 1.2$ and 
${\rm log}(f_{100}/f_{60})= 0.0 - 0.6$ 
(e.g., Sauvage, Thuan 1994). 
It is still possible that a few of the peculiar FIR objects are 
obscured galaxies with intrinsically low surface brightness.

\subsection
{Distances and Luminosities of the Peculiar FIR Objects}

About half of the 21 peculiar FIR objects are located within $b=2^\circ$ 
and 90\% are within $4^\circ$ (see table 2). This concentration 
toward the galactic plane is higher compared with that of globules 
by Clemens and Barvainis (1988), who found the extent of the
globules in $b$ to be $10^\circ$ to 12$^\circ$.   
This suggests that
the distances to the peculiar FIR objects are further than
the typical distances of dark globules, $\sim 600$ pc 
(Clemens, Barvainis 1988). If we take $\pm4^\circ$ as the extent in $b$
of the peculiar FIR objects and 87 pc as the layer thickness, the same
as that of molecular gas within 1 kpc by Dame et al. (1987), the typical
distances of the peculiar FIR objects are $\sim 1.2$ kpc. 
Note that the angular scale height of the ultra-compact H {\sc ii} regions
is $0.\!^\circ6$, which corresponds to $\sim$90 pc at the distance of the 
galactic center (Wood, Churchwell 1989).

For an assumed distance of 1 kpc, the peculiar FIR objects
have luminosities of $\sim$4 $L_\odot$ 
to 96 $L_\odot$ (see table 2), which correspond to the highest 
values of YSOs in the Taurus--Auriga and Ophiucus regions (Kenyon et al.
1990).  The peculiar FIR objects should have
internal energy sources, and they seem to have intermediate YSO luminosities.

The beam size, 21$''$, of Wouterloot and Brand$'$s (1989) observations 
at IRAM for the 19 objects, except for WB 901 and WB 1001, 
corresponds to about 0.1$d$ pc, where $d$ is an assumed distance in kpc, 
and is comparable to or smaller than the 
typical CO core sizes at distances of $d<$3 kpc. 
Thus, the weak CO intensities of the peculiar FIR objects
are not caused by dilution effects.

\subsection
{Location of the Peculiar FIR Objects}

The cirrus 2 flags (CIRR2) in IRAS PSC indicate a ratio of the cirrus flux,
$F_{\rm C}$ to source flux at 100 $\mu$m, $F_{\rm S}$:
\begin{equation}
CIRR2 = (8/3) \log (F_{\rm C}/F_{\rm S}) + (19/3).
\end{equation}
The distributions of the cirrus 2 flags are shown in table 4 
for the control sample and the peculiar 
FIR objects; the peculiar FIR objects have flags by two grades 
higher than the controle sample. The difference almost corresponds to the 
different 100 $\mu$m intensities between the peculiar FIR objects
and the control sample (see figures 1 and 3); the cirrus intensities are not
stronger around the peculiar FIR objects compared with the locations
of the control sample.
 
YSOs leave its parent molecular cores with time, and the
offset distances may become up to 1 pc for bright YSOs (e.g., 
Kenyon et al. 1990; Clark 1991). 
The peculiar FIR objects may have 
its parent molecular cores in the neighborhood. 
Dobashi et al. (1994) made 
a $^{13}$CO($J=1-0$) survey at a region of 
$\ell=80^\circ$ to $104^\circ$ and $b=-7.\!^\circ5$ to $10.\!^\circ5$
with $2.\!'7$ angular resolution and $8'$ grid spacing; 
they detected emission lines $>$1.2 K km s$^{-1}$ at 2191
positions, or 9\% of the observed points.
Yonekura et al. (1997) made a successive $^{13}$CO survey 
in the region $\ell=100^\circ$ to $130^\circ$ and 
$b=-10.\!^\circ5$ to $20^\circ$ with a same beam size and a same grid 
spacing as Dobashi et al. (1994).
Although there are 15 peculiar FIR objects among the regions of these
two surveys, 
none of them are associated with the molecular clouds. 
The peculiar FIR objects at other regions are also not located 
on any molecular clouds detected by Dame et al. (1987).

At $\ell<140^\circ$, 7 of the 17 peculiar FIR objects are located
on or near to large Lynds dark clouds; these are WB 92 vs. Lynds 1083,
WB 151 and 153 vs. Lynds 1107, WB 274 vs. Lynds 1238,
WB 302 vs. Lynds 1264, and WB 425 and 428 vs. Lynds 1373 (Lynds 1962).
These imply that the peculiar FIR objects tend to be associated 
with large dark clouds, or to be located at distances further than
the large dark clouds.
Yun and Clemens (1995) found 22 YSOs with near-infrared 
counterparts in globules and classified them into 10 class I 
objects and 12 class II objects. Their class I YSOs have $K$ = 9.8 to
13.9 and the luminosities at assumed distance of 600 pc are similar
to those of the peculiar FIR objects.
There is, however, an outstanding difference that Yun and Clemens'
class I YSOs almost have CO outflows (Yun, Clemens 1994), in contrast
with the weak CO emissions of the peculiar FIR objects.

Four peculiar FIR objects (WB 72, 92, 127, and 157) are located 
near to radio sources with 18 $-$ 39 mJy at $\lambda$ $\simeq$ 6 cm 
(Gregory et al. 1996); the separations between the FIR and 
radio sources are $38''$ to 108$''$.
None of the peculiar FIR objects are on the X-ray point sources detected 
by ROSAT (Voges et al. 1994).

\section
{Summary}

We found the presence of unusual IRAS point sources in the sense
that they are bright at 25, 60, and 100 $\mu$m but invisible
at optical and near-infrared wavelengths as well as in the $^{12}$CO line.
At least 18 such objects have $\alpha$-indices between 2.2 
and 25 $\mu$m, corresponding to class I YSOs.
The IRAS colors are similar to those of the usual YSOs; in detail,
about half of them have a second peak at between 2.2 $\mu$m and 25 $\mu$m
and the largest log($f_{100}/f_{60}$) values of YSOs.
Although the peculiar FIR objects avoid molecular clouds, 
seven objects are located
near to large dark clouds, and four are near to radio sources.
The distances are induced to be around 1 kpc, implying that 
they are mostly YSOs with intermediate luminosities and low CO
abundance in the envelopes. 

Further observations of these objects in molecular lines and 
millimeter continuum are needed to clarify their nature.

\par
\vspace{1pc} \par
SCANPI processing for IRAS data used in this study was made at IPAC (Infrared Processing and Analysis Center), which is operated by the California Institute of Technology, Jet Propulsion Laboratory under contract to the National Aeronautics and Space Administration (NASA). 
The Digitized Sky Surveys were produced at the Space Telescope Science Institute under  U.S. Government grant NAG W-2166. 
We thank an anonymous referee who gave us helpful comments.

\clearpage

\section*{References}
\small

\re
Acker A., Ochsenbein F., Stenholm B., Tylenda R., Marcout J.,
 Schohn C. 1992, Strasboug-ESO Catalogue of Galactic Planetary Nebulae
 (ESO, M\"unchen)

\re
Aikawa Y., Umebayashi T., Nakano T., Miyama S.M. 1997, ApJ 486, L51

\re
Benson P.J., Myers P.C. 1989, ApJS 71, 89

\re
Bessell M.S., Brett J.M. 1988, PASP 100, 1134

\re
Boulanger F., Baud B., van Albada G.D. 1985, A\&A 144, L9

\re
Boulanger F., Perault M., 1988, ApJ 330, 964

\re
Casoli F., Dupraz C., Gerin M., Combes F., Boulanger F., 1986, A\&A 169, 281

\re
Clark F.O. 1991, ApJS 75, 611

\re
Clemens D.P., Barvainis R. 1988, ApJS 68, 257

\re
Clemens D.P., Yun J.L., Heyer M.H. 1991, ApJS 75, 877

\re
Dame T.M., Ungerechts H., Cohen R.S., de Geus E.J., Grenier I.A., 
May J., Murphy D.C.,
Nyman L.-\AA., Thaddeus P. 1987, ApJ 322, 706

\re
Dobashi K., Bernard J.-P., Yonekura Y., Fukui Y. 1994, ApJS 95, 419

\re
Dutrey A., Guilloteau S., Duvert G., Prato L., Simon M.,
 Schuster K., M\'enard F. 1996, A\&A 309, 493

\re
Elias J.H., Frogel J.A., Matthews K., Neugebauer G. 1982, AJ 87, 1029

\re
Emerson J.P. 1988, in Formation and Evolution of Low Mass
Stars, ed A.K. Dupree, M.T.V.T. Lago (Kluwer, Dordrecht) p193

\re
Feigelson E.D. 1996, ApJ 468, 306

\re
Gregory P.C., Scott W.K., Douglas K., Condon J.J. 1996,
ApJS 103, 427

\re
Iwata I., Nakanishi K., Takeuchi T., Sait\=o M., Yamashita T.,
 Nishihara E., Okumura S. 1997, PASJ 49, 47

\re
Joint IRAS  Science Working Group 1988, IRAS Point Source Catalog,
version 2 (NASA, Washington D.C.) (IRAS PSC)

\re
Kenyon S.J., Hartmann L.W., Strom K.M., Strom S.E., 1990, AJ 99, 869

\re 
Lada C.J. 1987, in Star Forming Regions, ed M. Peimbert, 
J. Jugaku (Reidel, Dordrecht) p1

\re
Lynds B.T. 1962, ApJS 7, 1

\re
Mathis J.S. 1990, ARA\&A 28, 37

\re
Mermilliod J.-C., Mermilliod M. 1994, Catalogue of Mean UBV
 data on Stars (Springer-Verlag, New York)

\re
Nakanishi K., Takata T., Yamada T., Takeuchi T.T., Shiroya R.,
 Miyazawa M., Watanabe S., Sait\=o M. 1997, ApJS 112, 245

\re
Neuh\"auser R. 1997, Science 276, 1363

\re
Nilson P. 1973, Uppsala General Catalogue of Galaxies
 (Uppsala Offset Center, Uppsala)

\re
Sauvage M., Thuan T.X. 1994, ApJ 429, 153

\re
Smithsonian Astrophysical Observatory 1966, Star Catalog
 (Smithsonian Institution, Washington D.C.) (SAO)

\re 
Takata T., Yamada T., Sait\=o M., Chamaraux P., Kaz\'es I.
 1994, A\&AS 104, 529

\re
Voges W., Gruber R., Haberl F., Kuerster M., Pietsch W.,
 Zimmermann U. 1994, ROSAT Source Catalog, Version 11-May-1995

\re
Wackering L.R. 1970, Mem. RAS 73, 153

\re
Wainscoat R.J., Cowie L.L. 1992, AJ 103, 332

\re
Walker C.K., Adams F.C., Lada C.J. 1990, ApJ 349, 515

\re
Wood D.O.S., Churchwell E. 1989, ApJ 340, 265

\re
Wouterloot J.G.A., Brand J. 1989, A\&AS 80, 149

\re
Wouterloot J.G.A., Brand J., Burton W.B., Kwee K.K. 1990, A\&A 230, 21

\re
Yamada T., Takata T., Djamaluddin T., Tomita A., Aoki K.,
 Takeda A., Sait\=o M. 1993, ApJS 89, 57

\re
Yamashita T., Nishihara E., Okamura S., Mori A., Watanabe E. 1995, 
in Scientific and  Engineering Frontiers for 8--10 m Telescopes in 
the 21th Cettury, ed M. Iye, T. Nishimura 
(University Academy Press, Tokyo) p285

\re
Yonekura Y., Dobashi K., Mizuno A., Ogawa H., Fukui Y. 1997,
ApJS 110, 21

\re
Yun J.L., Clemens D.P. 1994, ApJS 92, 145

\re
Yun J.L., Clemens D.P. 1995, AJ 109, 742


\onecolumn


\begin{table}[t]
Table 1.\hspace{4pt}Optical counterparts of IRAS point sources not associated 
with the $^{12}$CO line in Wouterloot and Brand$'$s (1989) search.\\
\vspace{6pt}
\begin{tabular*}{\textwidth}{@{\hspace{\tabcolsep}
\extracolsep{\fill}}p{14pc}ll} \hline
Optical object & Number & Reference \\ \hline
star \dotfill  & 41 & SAO Star Catalog (1966), Wackering (1970), \\ 
                &    & Mermilliod, Mermilliod (1994) \\
galaxy$^*$ \dotfill & 79 & Nilson (1973), Yamada et al. (1993), \\
           &    &  Takata et al. (1994), Nakanishi et al. (1997) \\
planetary nebula \dotfill & 12 & Acker et al. (1992) \\
unknown  \dotfill  & 91 &   \\
\hline
\end{tabular*}
\vspace{6pt} \par\noindent
$*$ The number of galaxies with known redshifts is 65 among them.
\end{table}


\begin{table}[t]
\scriptsize  
\begin{center}
Table 2.\hspace{4pt}IRAS data of the point sources invisible at optical and 
near-infrared wavelengths \\
and having no detectable $^{12}$CO line. \\
\end{center}
\vspace{6pt}
\begin{tabular}{lllllllllllll} \hline

WB  & IRAS name &$\ell$& $b$& $f_{12}$ & $f_{25}$& $f_{60}$ & $f_{100}$ & 
\multicolumn{3}{c}{logarithm of} & $L_{IR}^{^*}$ & NIR \\

No. &           &  &  &(Jy)&(Jy)&(Jy)& (Jy)& 
$f_{25}/f_{12}$	& $f_{60}/f_{25}$ & $f_{100}/f_{60}$ & $(L_\odot)$ &
 obs.$^\dagger$ \\ \hline

  70&21188+5517& 96.05&   4.06&0.49 &0.49&10.38&43.63&    &    &    &  &   \\
    &          &      &       &0.55 &0.55&10.44&52.16&0.0 &1.28&0.70&48&1  \\
  72&21199+4949& 92.32&   0.06&0.25L&1.30& 7.70&15.39&    &    &    &  &   \\
    &          &      &       &0.07 &1.52& 8.79&15.41&1.34&0.76&0.24&25&1  \\
  92&21306+5532& 97.45&   3.09&0.73 &1.91&24.86&65.19&    &    &    &  &   \\
    &          &      &       &0.85 &2.37&30.33&76.67&0.45&1.11&0.40&96&1  \\ 
 117&21422+5625& 99.24&   2.67&0.36 &0.49& 6.94&31.35&    &    &    &  &   \\
    &          &      &       &0.51 &0.59& 6.63&38.02&0.06&1.05&0.76&35&1  \\
 119&21426+5723& 99.92&   3.38&0.25L&0.28& 3.39&12.66&    &    &    &  &   \\
    &          &      &       &0.27 &0.37& 3.50&12.47&0.14&0.98&0.55&14&1  \\
 127&21472+5642& 99.95&   2.44&0.25L&0.29& 4.22&18.15&    &    &    &  &   \\
    &          &      &       &0.22 &0.39& 5.22&15.73&0.25&1.13&0.48&18&1  \\
 151&22035+5442&100.56&$-$0.54&0.25L&0.30& 3.90& 8.13&    &    &    &  &   \\
    &          &      &       &0.11 &0.27& 4.01& 8.63&0.39&1.17&0.33&12&1  \\
 153&22044+5451&100.74&$-$0.49&0.25L&0.25& 1.45& 4.28&    &    &    &  &   \\
    &          &      &       & --- &0.26& 1.49& 6.97& ---&0.76&0.67&7:&1  \\
 157&22096+5725&102.81&   1.19&0.25L&0.40& 7.70&25.57&    &    &    &  &   \\
    &          &      &       &0.30 &0.51& 8.96&28.61&0.23&1.24&0.50&31&1  \\
 168&22169+5316&101.37&$-$2.83&0.25L&0.43& 1.31& 2.53&    &    &    &  &   \\
    &          &      &       &0.06 &0.27& 1.34& 2.87&0.65&0.70&0.33& 4&1  \\
 246&23055+5913&110.10&$-$0.77&0.26L&0.40& 6.87&27.07&    &    &    &  &   \\
    &          &      &       &0.18 &0.71& 6.95&30.83&0.60&0.99&0.65&29&1  \\
 260&23122+5758&110.45&$-$2.25&0.34 &0.96& 6.86&11.94&    &    &    &  &   \\
    &          &      &       &0.39 &1.82& 7.14&10.91&0.67&0.59&0.18&23&1  \\
 272&23190+5637&110.83&$-3.84$&0.25L&0.50& 1.65& 3.02&    &    &    &  &   \\
    &          &      &       &0.11 &0.54& 1.86& 3.34&0.69&0.54&0.25& 6&1  \\
 274&23217+6028&112.45&$-0.33$&0.37L&0.53& 4.26&14.79&    &    &    &  &   \\
    &          &      &       &0.11 &0.63& 4.20&17.53&0.76&0.82&0.62&18&1  \\
 302&23504+6802&117.55&   6.07&0.26 &0.52& 5.49&13.65&    &    &    &  &   \\
    &          &      &       &0.32 &0.58& 5.42&12.45&0.26&0.97&0.36&18&1  \\
 425&02264+6034&134.70&   0.20&0.44 &0.63&12.96&50.79&    &    &    &  &   \\
    &          &      &       &0.53 &0.95&14.65&58.94&0.25&1.19&0.60&59&1  \\
 428&02309+6034&135.21&   0.41&0.59 &0.96& 9.59&29.53&    &    &    &  &   \\
    &          &      &       &0.70 &1.18&11.32&35.91&0.23&0.98&0.50&43&1  \\
 584&04375+5016&155.50&   2.65&0.31L&0.47& 4.54& 8.57&    &    &    &  &   \\
    &          &      &       &0.37 &0.47& 4.53& --- &0.10&0.98&--- &10:&1 \\
 628&05183+3323&173.43&$-$1.86&0.66 &1.02&14.70&44.07&    &    &    &  &   \\
    &          &      &       &0.75 &1.29&12.78&56.27&0.24&1.00&0.64&57&H  \\
 901&06486-0110&214.02&$-$0.61&0.48 &0.54& 5.88&23.50&    &    &    &  &   \\
    &          &      &       &0.55 &0.61& 6.25&26.04&0.04&1.01&0.62&28&J  \\
1001&07272-1909&234.31&$-$0.68&0.25L&0.28& 2.31& 8.87&    &    &    &  &   \\
    &          &      &       &0.30 &0.37& 2.79&11.08&0.09&0.88&0.60&13&J  \\
\hline
\end{tabular}
\vspace{6pt} \par\noindent
$*$ For an assumed distance of 1 kpc.
\par\noindent
$\dagger$ 1: the exposure time was 300 s at $J$-band, 
150 s at $H$-band, and 225 s at $K'$-band. 
H: the exposure time was 600 s at $H$-band. 
J: the exposure time was 600 s at $J$-band.
\end{table}


\begin{table}[t]
\begin{center}
Table 3.\hspace{4pt}IRAS point sources without CO emission and associated 
with known stars.  \\
\end{center}
\vspace{6pt}
\begin{tabular*}{\textwidth}{@{\hspace{\tabcolsep}
\extracolsep{\fill}}lllllllllll} \hline

WB  & IRAS name &$\ell$& $b$& $f_{12}$ & $f_{25}$& $f_{60}$ & $f_{100}$ & 
Star & Sep.$^*$ & Ref.$^\dagger$ \\

No. &           &  &  &(Jy)&(Jy)&(Jy)& (Jy)& 
 SAO Sp $m_{\sc v}$ &  &  \\
\hline
17& 20556+4806& 88.34& 1.80& 0.35& 0.91& 9.75& 27.05& 050280 A0 8.4
& 6$''$& 1 \\
164& 22150+6109& 105.49& 3.89& 0.35& 1.92& 36.71& 78.53& Em. star 11B
& 2$''$& 2 \\
167& 22162+5539& 102.60& $-0.78$& 0.70& 2.49& 36.64& 93.46& 
 B0 10.4 & 15$''$& 3 \\
207& 22444+5853& 107.51& 0.09& 0.43& 1.57& 14.38& 33.75& Em. star 
& 35$''$& 2 \\
454& 02495+6414& 135.64& 4.64& 0.40& 1.15& 11.77& 26.01& 012493 
B8 8.6 & 9$''$& 1 \\
575& 04331+5211& 153.62& 3.40& 0.23& 0.46& 2.95& 7.62&  024716 A0 8.7
& 22$''$& 1  \\
600& 05017+2639& 176.80& $-$8.71& 0.43& 0.69& 2.48& 6.59&  076945 
A2 7.7 & 7$''$& 1 \\
645& 05247+3422& 173.37& $-$0.20& 1.66& 5.03& 56.89& 135.78& 058067
 B0 9.1 & 64$''$& 1 \\
655& 05293+1701& 188.50& $-$8.89& 0.43& 1.23& 3.74& 4.06 &  094630 
B9 6.0 & 4$''$& 1 \\
893& 06467$-$1505& 226.21& $-$7.38& 0.43& 1.62& 4.60& 5.68& 151962 
B5 5.3 & 15$''$& 1 \\
1177&08428$-$3657& 258.27& 3.52& 0.26& 0.88& 5.05& 9.21& 199573 B8 5.8
& 64$''$& 1 \\
\hline
\end{tabular*}
\vspace{6pt} \par\noindent
$*$ The separations given in IRAS PSC.
\par\noindent
$\dagger$ 1: Star Catalog (SAO 1966),
2: Wackering (1970), and 3: Mermilliod and Mermilliod (1994). 
\end{table}


\begin{table}[t]
\small
\begin{center}
Table 4. Distribution of IRAS cirrus 2 flags for the control sample
 and the peculiar FIR objects. 
\end{center}
\vspace{6pt}
\begin{tabular*}{\textwidth}{@{\hspace{\tabcolsep}
\extracolsep{\fill}}p{10pc}llllllllll} \hline
Flag  \dotfill & 1& 2 & 3& 4 & 5& 6& 7& 8& 9& total \\ \hline
Control sample \dotfill & 1& 3& 20& 21& 28& 19& 4& 1& 0& 97 \cr
Peculiar objects \dotfill & 0& 0& 1& 1& 2& 6& 7& 3& 1& 21 \cr
\hline
\end{tabular*}
\end{table}

\clearpage


\begin{figure}[ht]
\begin{center}
\psbox[width=160mm,vscale=1.0]{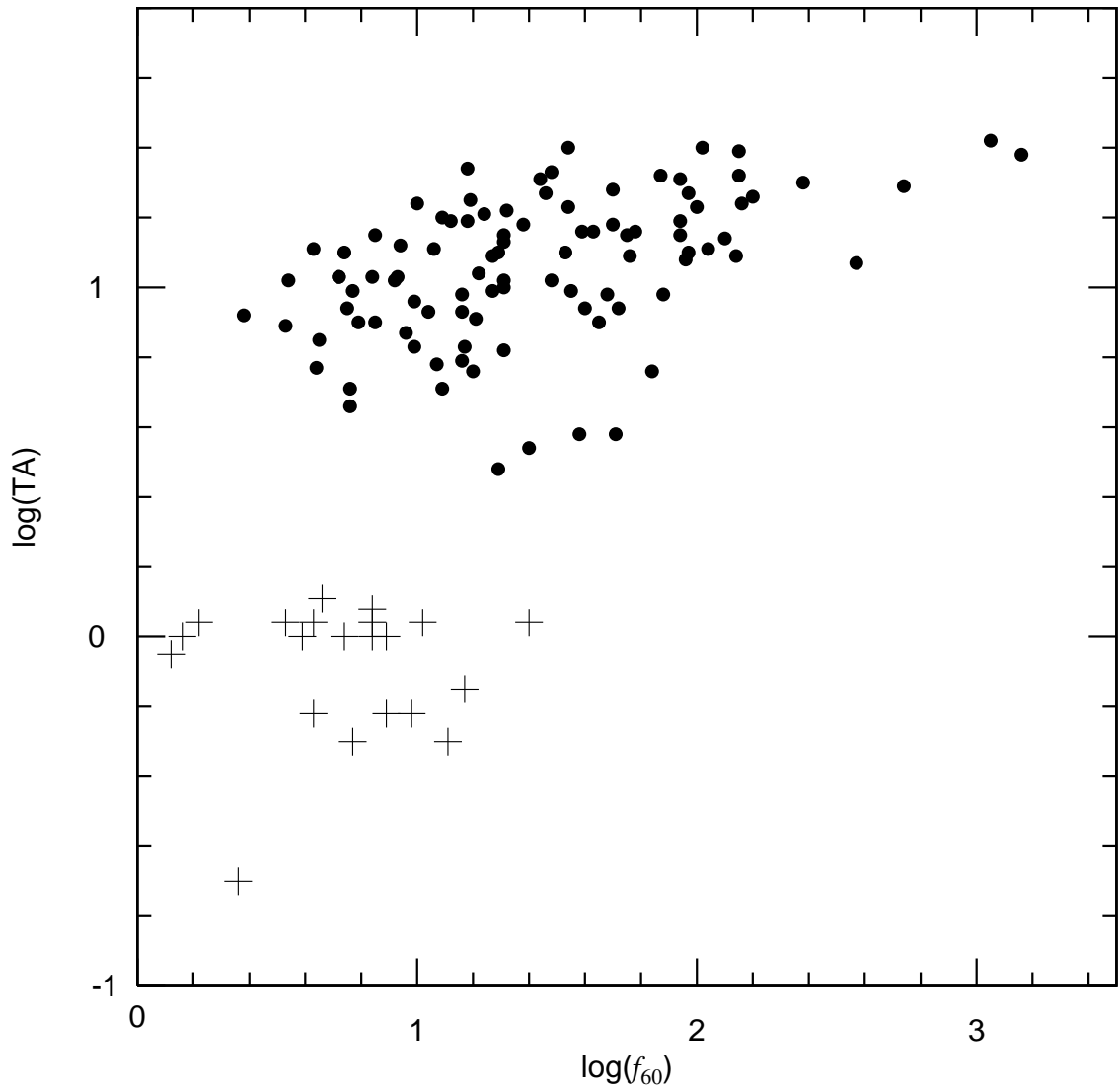}
\end{center}
\caption{
IRAS 60 $\mu$m flux density (Jy) versus the $^{12}$CO($J=1-0$) line
intensity (K) for IRAS point sources obtained by Wouterloot and Brand 
(1989). The dots indicate 97 objects with $^{12}$CO emission.
The crosses indicate 21 objects, the CO intensities of which are the upper
limits in the Wouterloot and Brand$'$ observations. 
The 21 objects are also invisible in the optical and near-infrared 
wavelengths.
}
\label{fig:1}
\end{figure}

\clearpage


\begin{figure}[ht]
\begin{center}
\fbox{\rule[-5cm]{0cm}{10cm}
\rule{5cm}{0cm}
see fig2[a,b,c].jpg files
\rule{5cm}{0cm}
}
\end{center}
\caption{
Sample images at optical and near-infrared wavelengths around
peculiar FIR objects. The optical images were made from the 
Dizitized Sky Survey 
and the near-infrared images were taken at OAO by using OASIS;
the 18 near-infrared images, except for last three, are at the $K'$ band, 
and for IRAS 05183+3323 it is at the $H$ band, and for the last two 
at the $J$ band. 
The FOV is $1.2'\times 1.2'$. The ellipses are the position uncertainty
ellipses taken from IRAS PSC.
}
\label{fig:2}
\end{figure}

\clearpage


\begin{figure}[ht]
\begin{center}
\psbox[width=80mm,vscale=1.0]{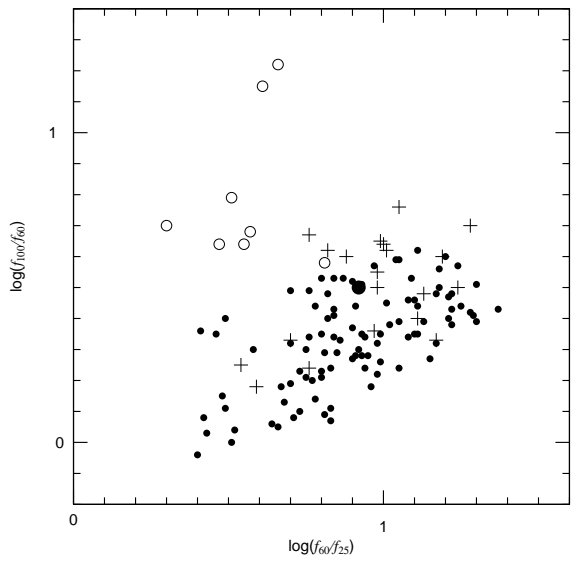}
\end{center}
\caption{
IRAS color--color diagram of the peculiar FIR 
objects (crosses), the control sample (dots), and
cirrus (open and filled circles).
$f_\lambda$ is the IRAS flux density at $\lambda$ $\mu$m.
}
\label{fig:3}
\end{figure}


\begin{figure}[ht]
\begin{center}
\psbox[width=80mm,vscale=1.0]{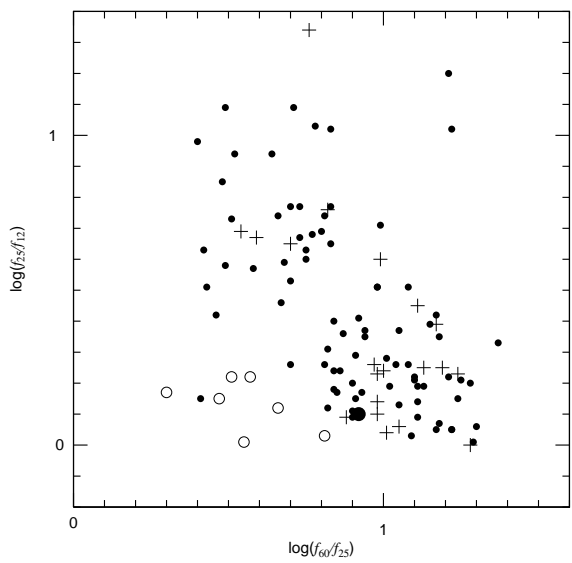}
\end{center}
\caption{
Same as figure 3, but for the colors of log($f_{60}/f_{25}$)
versus log($f_{25}/f_{12}$).
}
\label{fig:4}
\end{figure}

\label{last}
\end{document}